\documentclass[11pt, onecolumn]{article}
\usepackage{epsfig, psfrag}
\begin{document}

\title{An Addendum to ``How Good is PSK for Peak-Limited Fading Channels in the Low-SNR Regime?''}
\author{Wenyi Zhang, {\it Member, IEEE}
\thanks{The author is with Ming Hsieh Department of Electrical Engineering, University of Southern California, Los Angeles, CA. Email: {\tt wenyizha@usc.edu}. This work has been supported in part by NSF OCE-0520324.}
}
\date{}

\maketitle

\begin{abstract}
A proof is provided of the operational achievability of $R_\mathrm{rt}$ by the recursive training scheme in \cite{zhang07:it}, for general wide-sense stationary and ergodic Rayleigh fading processes.
\end{abstract}

\section{Introduction}

In \cite{zhang07:it}, an issue remaining open is whether the recursive training scheme interpretation \cite[Sec.~III.B]{zhang07:it} is operational for fading processes with an infinitely long memory length. As commented in \cite{zhang07:it}, the main difficulty in analysis pertains to the effect of error propagation, which cannot be readily circumvented in order to rigorously establish a channel coding theorem. In this addendum, we provide a proof of the operational achievability of $R_\mathrm{rt}$ by the recursive training scheme, for general wide-sense stationary and ergodic Rayleigh fading processes, without the $m$-dependence constraint as further imposed in the partial proof in \cite{zhang07:it}. The present proof exploits certain techniques in mismatched decoding, and its basic idea differs considerably from that in the previous proof for $m$-dependent fading processes.

Before proceeding to our proof, we briefly sketch a heuristic reasoning which has often been invoked to argue that the recursive training scheme is operational for channels with memory, and point out why such a heuristic reasoning is insufficient to lead to a mathematically rigorous proof. Consider the recursive training scheme as illustrated in \cite[Fig.~1]{zhang07:it}. The heuristic reasoning is as follows. For general fading processes that possess infinite memory, some residual correlation remains within each parallel sub-channel (PSC) among its $K$ symbols. Due to the ergodicity of the fading process, as we let the interleaving depth $L$ grow to infinity, this correlation asymptotically vanishes and thus each PSC can then be viewed as essentially memoryless. Meanwhile, as we let the coding blocklength $K$ grow to infinity, the channel coding theorem ({\it e.g.}, \cite{gallager68:book}) for the $L$ asymptotically memoryless PSCs ensures that each PSC can achieve a rate arbitrarily close to the input-constrained capacity of a true memoryless fading channel, and consequently the arbitrarily reliably decoded symbols can be used as essentially error-free training pilots for the recursive channel estimation/decoding procedure.

Unfortunately, the outlined heuristic reasoning is inherently flawed. It does not properly resolve a tension between channel memory decorrelation and error propagation. The essentially memoryless PSCs are obtained at the cost of indefinitely increasing the interleaving depth $L$. As $L$ increases, the block error probability of decoding each PSC is required to decrease inversely proportionally with $L$, in order to prevent catastrophic error propagation, and thus the coding block length $K$ is required to increase correspondingly. However, it is unclear how valid the ``essentially memoryless'' property of PSCs is as $K$ scales with $L$ asymptotically, because a formal mathematical characterization of such a property would involve a limiting argument that fixes a sufficiently large $K$ and subsequently lets $L$ grow large. At this point, we have observed a logic loop between the scaling of $K$ and $L$. So even if the heuristic reasoning were indeed correct, it lacks a mathematical confirmation, except for fading processes with finite memory, namely, the so-called $m$-dependent fading processes.

In our proof in this addendum, we break the logic loop by not requiring the PSCs be essentially memoryless, and hence relieve the tension between channel memory decorrelation and error propagation, since $K$ and $L$ do not need to scale with each other. However, without the essentially memoryless condition, we can no longer ignore the residual correlation within each PSC. Fortunately, the generalized mutual information (GMI) in the theory of mismatched decoding (see, {\it e.g.}, \cite{lapidoth02:it} and references therein) provides us an alternative way of establishing the achievability of information rates (at least) up to the input-constrained capacity of an associated memoryless fading channel, as will be elaborated shortly.

\section{Proof}

Consider the recursive training scheme as described in \cite[Sec.~III.B]{zhang07:it}. We seek to prove that, for an arbitrarily small decoding error rate $\delta > 0$ and an arbitrarily small overall rate loss factor $\lambda > 0$, there exists a code such that the recursive coding scheme achieves the rate $(1 - \lambda)\cdot R_\mathrm{rt}$ with an average decoding error probability no greater than $\delta$, where $R_\mathrm{rt}$ is defined by \cite[eqn.~(23)]{zhang07:it}.

For our purpose, we also consider a virtual system consisting of infinitely many PSCs. Among those PSCs, PSC $l$ is a Rayleigh fading channel with perfect receive channel state information (CSI), and with independent and identically distributed (i.i.d.) phase-shift keying (PSK) inputs of average SNR $\rho[l]$ as defined by \cite[Eqn.~(19)]{zhang07:it}. Furthermore, the fading process within each PSC is memoryless, and the noise samples are also i.i.d. zero-mean circular complex Gaussian. By such a construction, the $L$-average capacity, namely the average input-constrained capacity of the first $L$ PSCs in this virtual system approaches $R_\mathrm{rt}$, as $L$ grows asymptotically large (cf. \cite[Sec.~III.A]{zhang07:it}). In our proof, we hence fix a sufficiently large $L$ such that the $L$-average capacity of the virtual system exceeds $(1 - \lambda/2)\cdot R_\mathrm{rt}$.

In order to complete the achievability proof, it suffices to show that PSC $l$ in the recursive training scheme achieves at least the input-constrained capacity of memoryless PSC $l$ in the virtual system, for every $l$ from $0$ to $(L - 1)$. In light of \cite[Eqn.~(18)-(19)]{zhang07:it} and the one-step MMSE prediction in recursive training, we can write the channel equation of PSC $l$ as
\begin{eqnarray}
\label{eqn:PSC-l}
X[l; k] = \sqrt{\rho[l]} \cdot \hat{H}_\mathrm{d}[l; k] \cdot S[l; k] + \bar{Z}[l; k], \quad k = 1, \ldots, K,
\end{eqnarray}
where the index $[l; k]$ denotes the $k$th channel use of PSC $l$. Due to the wide-sense stationarity and ergodicity of the original fading process, both $\{\hat{H}_\mathrm{d}[l; :]\}$ and $\{\bar{Z}[l; :]\}$ sequences are wide-sense stationary and ergodic with respect to index $k$. In this addendum, we further normalize them to be of unit variance and thus $\rho[l]$ is the average SNR. However, due to the non-negligible residual correlation among channel uses within PSC $l$, in general neither $\{\hat{H}_\mathrm{d}[l; :]\}$ nor $\{\bar{Z}[l; :]\}$ sequence can be a circular complex Gaussian process. Fortunately, we have and shall utilize a ``locally'' Gaussian property of (\ref{eqn:PSC-l}). That is, for each $k$, $\hat{H}_\mathrm{d}[l; k]$ and $\bar{Z}[l; k]$ are jointly Gaussian and in fact independent, as shown in \cite[Sec.~III.A]{zhang07:it}.

Let us evaluate the GMI of (\ref{eqn:PSC-l}) following \cite[Sec.~III]{lapidoth02:it}. For a given codebook $\left\{s_m[l; :]\right\}_{m = 1}^M$ and a given realization of received channel outputs $x[l; :]$ along with channel fading process $\hat{h}_\mathrm{d}[l; :]$, a mismatched channel decoder chooses the message $m$ that minimizes the metric
\begin{eqnarray}
D(m) = \frac{1}{K} \sum_{k = 1}^K \left|x[l; k] - \sqrt{\rho[l]}\cdot \hat{h}_\mathrm{d}[l; k]\cdot s_m[l; k] \right|^2.
\end{eqnarray}
That is, the decoder treats the PSC as if it is were memoryless Rayleigh fading with i.i.d. circular complex Gaussian noise. Consider the ensemble of all codebooks i.i.d. generated by a prescribed PSK constellation, and without loss of generality assume the first message is selected with its corresponding codeword transmitted. First, due to ergodicity, the metric $D(1)$ converges almost surely to unity, since
\begin{eqnarray}
T &=& \lim_{K \rightarrow \infty} D(1)\nonumber\\
&=& \lim_{K \rightarrow \infty} \frac{1}{K} \sum_{k = 1}^K \left|\bar{Z}[l; k]\right|^2\nonumber\\
&=& \mathbf{E}\left\{\left|\bar{Z}[l; k]\right|^2\right\} = 1, \quad \mbox{a.s.}
\end{eqnarray}
The probability that an incorrect codeword accumulates a metric smaller than unity decays exponentially in $K$, and the GMI is just the exponent given by
\begin{eqnarray}
I_\mathrm{GMI} = \sup_{\mu < 0} (\mu - \Lambda(\mu)).
\end{eqnarray}
The limiting log-moment generating function $\Lambda(\mu) = \lim_{K \rightarrow \infty} \frac{1}{K}\Lambda_K(K \mu)$ is induced via
\begin{eqnarray}
\Lambda_K(\mu) = \log \mathbf{E}\left\{\left.e^{\mu D(m^\prime)}\right| X[l; :], \hat{H}_\mathrm{d}[l; :]\right\},
\end{eqnarray}
for $m^\prime > 1$. Because the inputs within each codeword are i.i.d. PSK symbols, conditioning upon $X[l; :]$ and $\hat{H}_\mathrm{d}[l; :]$, we have
\begin{eqnarray}
\label{eqn:temp-Lambda}
\frac{1}{K}\Lambda_K(K \mu) = \frac{1}{K} \sum_{k = 1}^K \log \mathbf{E}\left\{\left.e^{\mu \left|X[l;k] - \sqrt{\rho[l]}\cdot \hat{H}_\mathrm{d}[l;k]\cdot S\right|^2} \right|X[l;k], \hat{H}_\mathrm{d}[l;k]\right\},
\end{eqnarray}
where the expectation is with respect to the PSK constellation $S$. For concreteness, let us denote the PSK constellation by $S \in \left\{\theta_j\right\}_{j = 1}^J$ with each constellation point selected equiprobably. Hence we can further write (\ref{eqn:temp-Lambda}) as
\begin{eqnarray}
\frac{1}{K}\Lambda_K(K\mu) = \frac{1}{K}\sum_{k = 1}^K \log \sum_{j = 1}^J e^{\mu \left|X[l;k] - \sqrt{\rho[l]}\cdot \theta_j \cdot \hat{H}_\mathrm{d}[l;k]\right|^2} - \log J.
\end{eqnarray}
Due to ergodicity, for every $\mu < 0$ within a certain finite neighborhood of zero, $\Lambda(\mu)$ converges almost surely to an expectation, as
\begin{eqnarray}
\Lambda(\mu) &=& \lim_{K \rightarrow \infty} \frac{1}{K} \Lambda_K(K\mu)\nonumber\\
&=& \mathbf{E}\left\{\log \sum_{j = 1}^J e^{\mu \left|X - \sqrt{\rho[l]} \cdot \theta_j \cdot \hat{H}_\mathrm{d}\right|^2}\right\} - \log J,\quad \mbox{a.s.},
\end{eqnarray}
where the expectation is with respect to the fading coefficient $\hat{H}_\mathrm{d} \sim \mathcal{CN}(0, 1)$ and the channel output corresponding to the first message, $X = \sqrt{\rho[l]} \cdot \hat{H}_\mathrm{d} \cdot S_1 + \bar{Z}$ with $\bar{Z} \sim \mathcal{CN}(0, 1)$ independent of $\hat{H}_\mathrm{d}$. Note that the probability density function of $X$ conditioned upon $\hat{H}_\mathrm{d}$ is
\begin{eqnarray}
p_{X|\hat{H}_\mathrm{d}}(x|\hat{h}_\mathrm{d}) = \frac{1}{J} \sum_{j = 1}^J \frac{1}{\pi} e^{-\left|x - \sqrt{\rho[l]}\cdot\theta_j\cdot\hat{h}_\mathrm{d}\right|^2}.
\end{eqnarray}
So by letting $\mu = -1$, we have
\begin{eqnarray}
I_\mathrm{GMI} &\geq& -1 - \Lambda(-1)\nonumber\\
&=& -1-\log\pi - \mathbf{E}\left\{\left.
\mathbf{E}\left\{\log p_{X|\hat{H}_\mathrm{d}}(X|\hat{H}_\mathrm{d})\right\}
\right| \hat{H}_\mathrm{d}\right\}\nonumber\\
&=& h(X|\hat{H}_\mathrm{d}) - h(\bar{Z})\nonumber\\
&=& h(X|\hat{H}_\mathrm{d}) - h(X|S_1, \hat{H}_\mathrm{d}) = I(S_1; X|\hat{H}_\mathrm{d}),
\end{eqnarray}
which is precisely the input-constrained capacity of memoryless PSC $l$ in the virtual system. Hence as we choose a sufficiently large coding blocklength $K$ such that for every $l$ from $0$ to $(L - 1)$, there exist a codebook and an associated mismatched channel decoder that achieve a fraction of $(1 - \lambda)/(1 - \lambda/2)$ of the GMI of PSC $l$ with an average decoding error probability no greater than $\delta/L$, the entire recursive training scheme achieves the rate $(1 - \lambda)\cdot R_\mathrm{rt}$ with an average decoding error probability no greater than $\delta$, following a standard union upper bounding argument. This concludes our proof.

\bibliographystyle{unsrt}

\begin{thebibliography}{00}

\bibitem{zhang07:it} W. Zhang and J.~N. Laneman, ``How Good is PSK for Peak-Limited Fading Channels in the Low-SNR Regime?'' \textit{IEEE Trans. Inform. Theory}, Vol. 53, No. 1, pp. 236--251, Jan. 2007.

\bibitem{gallager68:book} R.~G. Gallager, \textit{Information Theory and Reliable Communication}, John Wiley \& Sons, Inc., New York, NY, 1968.

%
\bibitem{lapidoth02:it} A. Lapidoth and S. Shamai (Shitz), ``Fading Channels: How Perfect Need `Perfect Side Information' Be?'' \textit{IEEE Trans. Inform. Theory}, Vol. 48, No. 5, pp. 1118--1134, May 2002.

\end{thebibliography}

\end{document}